\documentclass[prd,preprint]{revtex4}
\usepackage{amsfonts}
\usepackage{amssymb}
\usepackage{amsmath}
\usepackage{graphicx}
\usepackage{graphics}
\usepackage{color}
\usepackage{epsfig}
\newcommand{\be}{\begin{equation}}
\newcommand{\ee}{\end{equation}}
\newcommand{\bea}{\begin{eqnarray}}
\newcommand{\eea}{\end{eqnarray}}
\newcommand{\sptwo}{1.4}
\newcommand{\doublespace}{\edef\baselinestretch{\sptwo}\Large\normalsize}
\newcommand{\newsection}[1]{\setcounter{equation}{0}}

\newcounter{newapp}
\begin{document}
\begin{titlepage}

\begin{center}
{\bf Transition Probability for Class of Two Level System Time Dependent Hamiltonians}
\end{center}
\begin{center}
{S.T. Love*}\\
{Department of Physics and Astronomy,\\
	Purdue University,\\
	West Lafayette, IN 47907-2036, U.S.A.}
\end{center}
\vspace{1in}
\begin{center}
{\bf Abstract}\\
Transition probabilities for a class of two level systems described by explicitly time dependent Hamiltonians are considered. Provided only that the approach to the infinite time limit is non-trivial  falling at least as fast as $1/ t$ for large $t$, the transition probability takes a particularly simple form depending only on the value of Hamiltonian parameters in this limit. 
\end{center}

\vspace{4in}
*e-mail address: loves@purdue.edu

\end{titlepage}
%\pagebreak

\doublespace

\underline\\

\noindent 
Among the simplest quantum mechanical Hamiltonians are those involving two-level systems. Various of these models have produced exactly soluble transition probabilities and have proven to be extremely illustrative and useful. Among the most celebrated examples of this type are the Landau-Zener model\cite{LL}-\cite{EM}, the Rabi model\cite{IR}, and its generalization, the Jaynes-Cummings model \cite{JC} as well as various soluble pulses\cite{RZ}-\cite{GDZ} and a few oscillatory models\cite{GDZ}-\cite{XH}. More recently, Barnes and Das Sarma\cite{BS}-\cite{MN} have introduced a method of reverse engineering the Schr\"odinger equation to deduce a large class of analytically soluble time dependent Hamiltonian driving terms.

In the present note, we consider a class of two level systems whose dynamics is governed by the Hamiltonian 
\bea 
H(t)&=&\frac{\hbar\Upsilon(t)}{2}  \sigma_z +\frac{\hbar\omega}{2} \sigma_x
\eea
where $\omega=\omega^* >0$ and $\sigma_i$ are the Pauli matrices. The real time dependent Hamiltonian function, $\Upsilon(t)=\Upsilon^*(t)$, can have an arbitrary time dependence restricted only by its exhibiting a non-trivial asymptotic in time dependence as it approaches the limit $\Upsilon(\infty)$ so that
\bea
\Upsilon(t)\sim \Upsilon(\infty)\left(1+{\cal O}\left(\frac{1}{\omega t}\right)\right)
\eea
That is, for $\omega t >>1$, $\Upsilon(t)$ approaches $\Upsilon(\infty)$ least as fast as $1/\omega t$, but does have some  non-trivial $t$ dependence. \\

For such Hamiltonians, we shall demonstrate that for a system prepared at time $t=0$ in a $\sigma_x$ eigenstate with eigenvalue $+1$, the transition probability to the state which is a $\sigma_x$ eigenstate with eigenvalue $-1$ is 
\bea
{\cal P}_{+\rightarrow -}(\infty)&=&\frac{1}{2}-\frac{1}{2}\frac{\omega}{\sqrt{\omega^2+\Upsilon^2(\infty)}}
\eea
This simple result holds for the entire class of Hamiltonians described above independent of the time dependence at non-infinite times. \\

To establish this claim, we begin by expanding the two level system state vector, $|\psi(t)>$, at time $t$ in terms of the $\sigma_x$ eigenstates, $|\pm>$ , satisfying
\bea
\sigma_x |\pm> &=&\pm |\pm>
\eea 
as
\bea
|\psi(t)>&=&\cos\Theta(t) e^{i\varphi_+(t)} e^{-i\frac{\omega t}{2}}|+> +\sin\Theta(t) e^{i\varphi_-(t)} e^{i\frac{\omega t}{2}}|-> 
\label{exp}
\eea
with $\Theta(t)=\Theta^*(t)$ and  $\varphi_\pm(t)=\varphi_\pm^*(t)$.

At time $t=0$, the system is prepared in the state $|+>$ so that 
\bea 
\cos\Theta(0)&=&1 \cr
\sin\Theta(0)&=&0
\eea
We seek the transition  probability to the state $|->$ which is given by
\bea
{\cal P}_{+\rightarrow -} (\infty)&=&\sin^2\Theta(\infty )
\eea
for the class of $\Upsilon(t)$ functions having the asymptotically approach to  $\Upsilon(\infty)$ detailed above. 

Application of the Schr\"odingr equation $i\hbar \frac{d}{dt}|\psi(t)> =H(t)|\psi(t)>$ to the above expansion (\ref{exp}) identifies
\bea
\Upsilon(t)&=&2i \left(-\frac{d\Theta(t)}{dt}+i\frac{d\varphi_+(t)}{dt}\frac{\cos\Theta(t)}{\sin\Theta(t)}\right)e^{i\varphi(t)}
\label{Upsilon1}
\eea
and 
\bea
\Upsilon(t)&=&2i \left(\frac{d\Theta(t)}{dt}+i\frac{d\varphi_-(t)}{dt}\frac{\sin\Theta(t)}{\cos\Theta(t)}\right)e^{-i\varphi(t)}
\label{Upsilon2}
\eea
where we have defined 
\bea
\varphi (t)&\equiv & \varphi_+(t)-\varphi_-(t) -\omega t
\label{varphi}
\eea
Equating the two expressions for $\Upsilon(t)$ then gives
\bea
2\cos\varphi(t)\frac{d\Theta(t)}{dt}&=&i\frac{d\varphi_+(t)}{dt}\frac{\cos\Theta(t)}{\sin\Theta(t)}e^{i\varphi(t)}-i\frac{d\varphi_-(t)}{dt}\frac{\sin\Theta(t)}{\cos\Theta(t)}e^{-i\varphi(t)}
\label{dotTheta}
\eea
Imposing the reality  of the time dependent driving term  $\Upsilon(t)=\Upsilon^*(t)$, then dictates 
\bea
\frac{d\varphi_+(t)}{dt}\frac{\cos\Theta(t)}{\sin\Theta(t)}&=&\frac{d\varphi_-(t)}{dt}\frac{\sin\Theta(t)}{\cos\Theta(t)}
\label{dotvarphipm}
\eea
or 
\bea
\frac{d\varphi(t)}{dt} +\omega &=& -\frac{d\varphi_-(t)}{dt}\frac{\cos(2\Theta(t))}{\cos^2\Theta(t)}
\label{dotvarphi}
\eea
Substituting Eq. (\ref{dotvarphipm}) into Eq. (\ref{dotTheta}) gives
\bea
\frac{d\Theta(t)}{dt}&=&-\frac{d\varphi_-(t)}{dt}\frac{\sin\Theta(t)}{\cos\Theta(t)}\frac{\sin\varphi(t)}
{\cos\varphi(t)}
\label{dotTheta1}
\eea
which when substituted into Eq. (\ref{Upsilon2}) yields
\bea
\Upsilon(t)&=&-2\frac{d\varphi_-(t)}{dt}\frac{\sin\Theta(t)}{\cos\Theta(t)}\frac{1}{\cos\varphi(t)}
%\label{Upsilon3}
\eea
or after using Eq. (\ref{dotTheta1}) that
\bea
\Upsilon(t) &=&\frac{2}{\sin\varphi(t)}\frac{d\Theta(t)}{dt}
\label{Upsilon4}
\eea
Substituting Eq. (\ref{dotvarphi}) into Eq. (\ref{dotTheta1}) relates $\Theta(t)$ and $\varphi(t)$ as
\bea
2\frac{d\Theta(t)}{dt}&=& \left(\frac{d\varphi(t)}{dt}+\omega\right)\frac{\sin\varphi(t)}{\cos\varphi(t)}
\eea
which can be integrated to give
\bea
\sin(2\Theta(t))\cos\varphi(t) &=& \Phi(t)
\label{2Theta}
\eea
where  
\bea
\Phi(t)&=& e^{\omega \int dt \tan\varphi(t)}
\eea
so that 
\bea
ln~ \Phi(t) =\omega \int dt \tan\varphi(t)
\eea
Note that since the left side of Eq. (\ref{2Theta}) has magnitude less than or equal to one, it follows that  $\omega\int dt \tan\varphi(t) \le 0$ and as such  $\Phi(t)$ is constrained as
\bea
0\le \Phi(t)< 1
\label{constraint1}
\eea
Using that
\bea
\tan\varphi(t)&=&\frac{1}{\omega}\frac{d}{dt}ln~\Phi(t)=\frac{1}{\omega}\frac{1}{\Phi(t)}\frac{d\Phi(t)}{dt}
\eea
Eq. (\ref{2Theta}) can be cast as
\bea
\sin(2\Theta(t))&=&\sqrt{\Phi^2(t)+\frac{1}{\omega^2}\left(\frac{d\Phi(t)}{dt}\right)^2}
\label{2Theta2}
\eea
from which follows the constraint
\bea
\Phi^2(t)+\frac{1}{\omega^2}\left(\frac{d\Phi(t)}{dt}\right)^2 <1
\label{constraint2}
\eea
which is consistent with the constraint of Eq. (\ref{constraint1}).\\
Note that the initial conditions
\bea
\cos\Theta(0)&=&1 \cr
\sin\Theta(0)&=&0
\eea
translate to 
\bea\Phi(0)&=&0 \cr
\frac{d\Phi(t)}{dt}|_{t=0}&=&0
\eea
Differentiating Eq.(\ref{2Theta2}) with respect to time and substituting for $\frac{d\Theta(t)}{dt}$ in Eq. (\ref{Upsilon4}) relates  $\Upsilon(t)$ to $\Phi(t)$ as
\bea
\Upsilon(t)&=&\frac{1}{\omega}\frac{\frac{d^2\Phi(t)}{dt^2}+\omega^2\Phi(t)}{\sqrt{1-\Phi^2(t)-\frac{1}{\omega^2}\left(\frac{d\Phi(t)}{dt}\right)^2}}
\label{Upsilon5}
\eea
Eq. (\ref{2Theta2}) can be recast as 
\bea
\cos^2\Theta(t)&=&\frac{1}{2}+\frac{1}{2}\sqrt{1-\Phi^2(t)-\frac{1}{\omega^2}\left(\frac{d\Phi(t)}{dt}\right)^2}\cr
\sin^2\Theta(t)&=&\frac{1}{2}-\frac{1}{2}\sqrt{1-\Phi^2(t)-\frac{1}{\omega^2}\left(\frac{d\Phi(t)}{dt}\right)^2}
\label{probs}
\eea
which incorporates the initial conditions.

In principle, given a form for $\Upsilon(t)$, one would solve the non-linear differential equation (\ref{Upsilon5}) for $\Phi(t)$ which when substituted into Eq. (\ref{probs}) gives the transition probability. Unfortunately, solving the differential equation is, in general, a highly formidable task. An alternate approach is to reverse engineer the problem\cite{BS} and consider various functions $\Phi(t)$ satisfying $\Phi(0)=\frac{d\Phi(t)}{dt}|_{t=0}=0$ and consistent with the constraints (\ref{constraint1}) and (\ref{constraint2}). Then using Eq. (\ref{Upsilon5}), the associated function $\Upsilon(t)$ is secured, while Eq. (\ref{probs}) gives the desired transition probability. The appendix details several examples of different choices for $\Phi(t)$, the corresponding Hamiltonian function $\Upsilon(t)$ and the resulting transition probability at time $t$.

Rather than focusing on specific choices for $\Phi(t)$ and/or $\Upsilon(t)$, let us consider the class of $\Upsilon(t)$ functions each of which has the property that it asymptotically approaches $\Upsilon(\infty)$ as
\bea
\Upsilon(t)\sim \Upsilon(\infty)\left(1+{\cal O}\left(\frac{1}{\omega t}\right)\right)
\eea
That is, for $\omega t >>1$, $\Upsilon(\infty)$ falls at least as fast as $1/\omega t$, but does have some  non-trivial $t$ dependence. Note that each of functions in the three examples in the appendix exhibit this behavior. On the other hand, this excludes the case where $\Upsilon(t)$ is $t$ independent for asymptotically large $\omega t$ in which case the transition probability varies sinusoidally at large $\omega t$ so that the $t\rightarrow \infty$ limit is ill defined.\footnote{For the specific case of the Rabi Hamiltonian\cite{IR}, where $\Upsilon(t)=\Upsilon$ is time independent for all $t$, it is straightforward to check that Eq. (\ref{Upsilon5}) is satisfied with $\Phi(t)=\frac{2\omega\Upsilon}{\omega^2+\Upsilon^2}\sin^2\left(\frac{\sqrt{\omega^2+\Upsilon^2}}{2}t\right)$. This, in turn, gives the well known transition probability at time $t$ as $\sin^2\Theta(t)=\frac{\Upsilon^2}{\omega^2 +\Upsilon^2}\sin^2\left(\frac{\sqrt{\omega^2+\Upsilon^2}}{2}t\right)$. Note that such a time independent driving term does not meet the criteria required for the current study. Moreover, 
since the probability has the purely sinusoidal behavior, it is ill defined in the limit $t\rightarrow \infty$.}  
This asymptotic approach of $\Upsilon(t)$ to $\Upsilon(\infty)$ then prescribes (c.f. Eq. (\ref{Upsilon5})) that  $\frac{d\Upsilon(t)}{dt}|_{t\rightarrow \infty}=\frac{d^2\Upsilon(t)}{dt^2}|_{t\rightarrow \infty}=0$ which in turn dictates that
\bea
\Phi(\infty)&=&\frac{\Upsilon(\infty)}{\sqrt{\omega^2+\Upsilon^2(\infty)}} \Leftrightarrow \Upsilon(\infty)=\omega \frac{\Phi(\infty)}{\sqrt{1-\Phi^2(\infty)}} 
\eea
while $\frac{d\Phi(t)}{dt}|_{t\rightarrow \infty}=
\frac{d^2\Phi(t)}{dt^2}|_{t\rightarrow \infty}=0$. \\
As such, the transition probability for all functions $\Upsilon(t)$ in this class takes the simple form
\bea
{\cal P}_{+\rightarrow -}(\infty)&=&\sin^2\Theta(\infty)\cr
 &=&  \frac{1}{2}-\frac{1}{2}\frac{\omega}{\sqrt{\omega^2+\Upsilon^2(\infty)}}
\label{main}
\eea
which is the claimed result. Note that this probability depends only on the parameters of the Hamiltonian at $t \rightarrow \infty$, namely $\omega$ and $\Upsilon(\infty)$  while being independent of the function form of $H(t)$ for all earlier times. We see that ${\cal P}_{+\rightarrow -}(\infty)$ increases monotonically as a function of $\left(\frac{\Upsilon(\infty)}{\omega}\right)^2$ from $0$ achieving its maximum value of $\frac{1}{2}$ as $\left(\frac{\Upsilon(\infty)}{\omega}\right)^2$ increases from $0$ to $\infty$.\\
\\

\noindent 
During the course of this study, I have enjoyed several useful conversations with T.E. Clark who I also thank for his help with the plotting.\\

\noindent 
{\bf Appendix: Reversed Engineered Time Dependent Hamiltonian Driving Terms} \\
\\
In this appendix, we provide three examples of functions $\Phi(t)$ all of which are consistent with the constraints of Eqs. (\ref{constraint1}) and (\ref{constraint2}) while satisfying $\Phi(0)=0$ and $\frac{d\Phi(t)}{dt}|_{t=0}=0$. On the other hand, the three functions have different time dependences for $t>0$. The Hamiltonian functions $\Upsilon(t)$ producing these various $\Phi(t)$ functions are then constructed using Eq. (\ref{Upsilon5}) and again display diverse behaviors at finite times. In each case, said functions exhibit a nontrivial asymptotic approach to $\Upsilon(\infty)$ given by $\Upsilon(t)\sim \Upsilon(\infty)+{\cal O}(\frac{1}{\omega t})$. The various transition probabilities, ${\cal P}_{+\rightarrow -}(t)$, are secured for all times $t>0$. While the three transition probabilities exhibit quite distinct behaviors at finite times, in the limit, $t\rightarrow \infty$, each of the transition probabilities takes the same form as given in Eq. (\ref{main}):
\bea
{\cal P}_{+\rightarrow -}(\infty) &=&\frac{1}{2}-\frac{1}{2}\frac{\omega}{\sqrt{\omega^2 +\Upsilon^2(\infty)}}
\eea
depending only on the parameters appearing in $\Upsilon(\infty)$ which exemplifies the general result presented in this paper. 
\\
Example 1: First consider 
\bea
\Phi(t)&=&\Phi(\infty) \left(\frac{\omega t}{\gamma^2 +\omega t}\right)^\eta
\eea
with $0\le \Phi(\infty) <1~~$, $\gamma \ge 1$  and $ \eta \ge 2$. Note that for this class of functions,  $\Phi(0)=\frac{d\Phi(t)}{dt}|_{t=0}=0$ while $\frac{d^2\Phi(t)}{dt^2}|_{t=0}=\frac{2\beta\omega^2}{\gamma^4}\Phi(\infty)$ for $\eta =2$ while vanishing for $\eta >2$.\\
Asymptotically
\bea
\Phi(t)&\sim &\Phi(\infty)\left(1-\frac{\eta \gamma^2}{\omega t}\right)
\eea
which approaches the value $\Phi(\infty)$ up to  terms of ${\cal O}(\frac{1}{\omega t})$, while $\frac{d\Phi(t)}{dt} $ and $\frac{d^2\Phi(t)}{dt^2}$ asymptotically vanish as $\frac{1}{(\omega t)^2}$ and $\frac{1}{(\omega t)^3}$ respectively. 

The Hamiltonian term $\Upsilon(t)$ producing such a $\Phi(t)$ is given by (c.f. Eq. (\ref{Upsilon5}))
\bea
\Upsilon(t)&=& \omega \frac{N_1(\omega t)}{D_1(\omega t)}
\eea 
where 
\bea 
N_1(\omega t)
&=&\left(1+\frac{\eta\gamma^2((\eta-1)\gamma^2-2\omega t)}{(\omega t)^2(\gamma^2+\omega t)^2}\right)\Phi(t)\cr
&=&\Phi(\infty)\left(1+\frac{\eta\gamma^2((\eta-1)\gamma^2-2\omega t)}{(\omega t)^2(\gamma^2+\omega t)^2}\right) \left(\frac{\omega t}{\gamma^2 +\omega t}\right)^\eta
\eea
and 
\bea 
D_1(\omega t)&=&\sqrt{1-\left(1+\frac{\eta^2 \gamma^4}{(\omega t)^2 (\gamma^2 +\omega t)^2}\right)\Phi^2(t)}\cr
&=&\sqrt{1-\Phi^2(\infty)\left(1+\frac{\eta^2 \gamma^4}{(\omega t)^2 (\gamma^2 +\omega t)^2}\right) \left(\frac{\omega t}{\gamma^2 +\omega t}\right)^{2\eta}}
\eea 
The ratio $\frac{\Upsilon(t)}{\omega}$ with parameters $\eta=2, \gamma  =1$ and $\Phi(\infty) =\frac{2\sqrt{2}}{3}$ is displayed in the figure as a function of $\omega t$ for  $0\le \omega t\le 700$. After an initial falloff (see the insert) from $\Upsilon(0)=\frac{4}{3}\sqrt{2}\omega$ to a minimum at $\omega t \simeq 0.75$, it smoothly rises  
approaching the finite value
\bea
\Upsilon(\infty)&=&\omega\frac{\Phi(\infty)}{\sqrt{1-\Phi^2(\infty)}}=\sqrt{8}\omega
\eea
with non-trivial sub-leading corrections varying as ${\cal O}(1/t)$ for large $t$. Moreover, $\frac{d\Upsilon(t)}{dt}$ and $\frac{d^2\Upsilon(t)}{dt^2}$ each approach zero asymptotically as $\frac{1}{(\omega t)^2}$ and  $\frac{1}{(\omega t)^3}$ respectively. 
\\

\begin{figure*}
	\begin{center}
		$\begin{array}{cc}
		\includegraphics[scale=0.4]{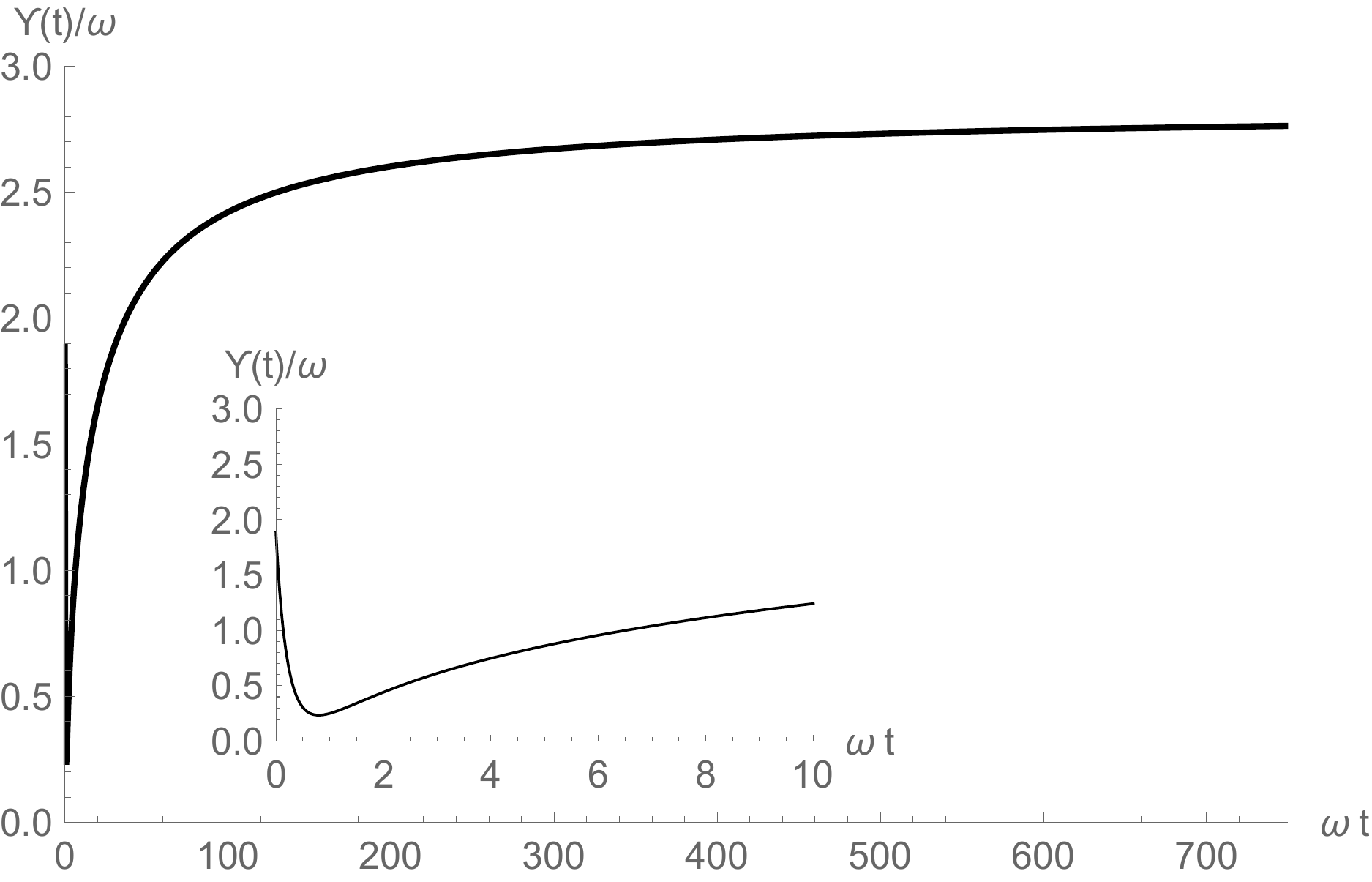}\hspace{.4in}\includegraphics[scale=0.4]{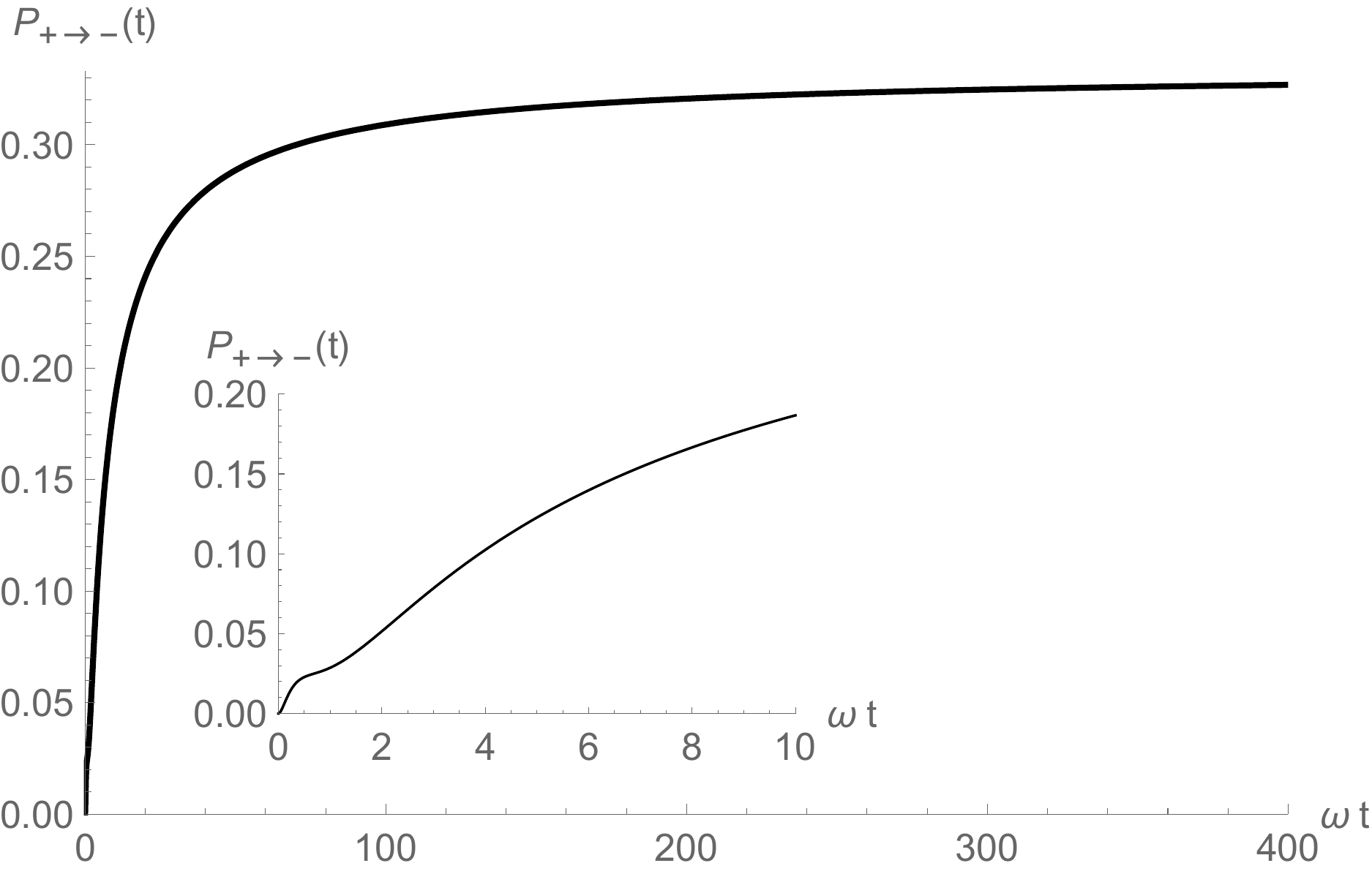}\\
		\end{array}$
		\caption{Example 1 with parameters $\gamma =1~;~\eta =2$ and $\Phi(\infty)=\frac{2\sqrt{2}}{3}$. \\
			Left panel: $\Upsilon(t)/\omega$ as a function of $\omega t$ with insert highlighting small $\omega t$ range;\\
			Right panel: transition probability ${\cal P}_{+\rightarrow -}(t)$ as function of $\omega t$ with insert highlighting small $\omega t$ range.} 
		\label{fig1}
	\end{center}
\end{figure*}

The transition probability at time $t$  is (c.f. Eq. (\ref{probs}))
\bea
{\cal  P}_{+\rightarrow-}(t)&=&\frac{1}{2}-\frac{1}{2}D_1(\omega t)
\eea
This probability is plotted in the figure as a function of $\omega t$ for $0 \le \omega t \le 400$. Said probability smoothly rises from the initial condition ${\cal P}_{+\rightarrow -}(0)=0$ to the value 
\bea
{\cal  P}_{+\rightarrow-}(\infty)&=&\frac{1}{2}-\frac{1}{2}\frac{\omega}{\sqrt{\omega^2+\Upsilon^2(\infty)}}|_{\frac{\Upsilon(\infty)}{\omega}=\sqrt{8}}=1/3
\eea
as $t\rightarrow \infty$ which has the form of the general result discussed in the body of the paper. Note that this probability depends only on the parameters of the Hamiltonian at $t\rightarrow \infty$, namely, $\omega$ and $\Upsilon(\infty)$, and is independent of the parameters at finite times, i.e. $\eta$ and $\gamma$.\\
\\
Example 2: Next consider factor
\bea
\Phi(t)&=& \Phi(\infty) e^{-\frac{\gamma}{\omega t}}
\eea 
with $0\le \Phi(\infty) <1$ and $\gamma \ge 1$. Comparing with the previous example, we see that the asymptotic in $t$ behavior is of a  similar nature for both functions going as $\Phi(t)=\Phi(\infty) +{\cal O}(\frac{1}{\omega t})$. On the other hand, for small $t$, the $\Phi(t)$ in this example exhibits an essential singularity at $t=0$, so that $\Phi(t)$ and all time derivatives of $\Phi(t)$ vanish at $t=0$, while in the previous example, $\Phi(t)\simeq t^\eta~~;~~t\rightarrow 0$ with $\eta \ge 2$.

The Hamiltonian term producing such a phase factor is readily secured as
\bea
\Upsilon(t)&=&\omega \frac{N_2(\omega t)}{D_2(\omega t)}
\eea
where
\bea
N_2(\omega t)
&=&\Phi(\infty) \left(\frac{\gamma^2}{(\omega t)^4}-\frac{2\gamma}{(\omega t)^3}+1\right)e^{-\frac{\gamma}{\omega t}}
\eea
and 
\bea
D_2(\omega t)&=&{\sqrt{1-\Phi^2(\infty)\left(1+\frac{\gamma^2}{(\omega t)^4}\right) e^{-\frac{2\gamma}{\omega t}}}}
\eea
The ratio $\frac{\Upsilon(t)}{\omega}$ is displayed in the figure as a function of $\omega t$ for $0 \le \omega t \le 100$ and with parameters $\gamma =1$ and $\Phi(\infty) =\frac{2\sqrt{2}}{3}$. Rising from zero at $t=0$,  the function exhibits  a sharp spike at $\omega t \simeq 0.25$ to a height of around 2.5 and before decreases to a minimum which is slightly negative at $\omega  t \sim 0.75$. This is seen in the embedded plot in the figure which highlights $\Upsilon(t)$ for the range $0 \le 0 \le 2$.  For larger values of $\gamma$, the peak broadens.  Subsequently, the function increases monotonically  again approaching the finite limit
\bea
\Upsilon(\infty)&=&\omega\frac{\Phi(\infty)}{\sqrt{1-\Phi^2(\infty)}}=\sqrt{8}\omega
\eea
while exhibiting the desired large $t$ behavior
\bea 
\Upsilon(t) &\sim & \omega \frac{\Phi(\infty)}{\sqrt{1-\Phi^2(\infty)}}\left(1-\frac{1}{1-\Phi^2(\infty)}\frac{\gamma}{\omega t}\right)
\eea

\begin{figure*}
	\begin{center}
		$\begin{array}{cc}
		\includegraphics[scale=0.4]{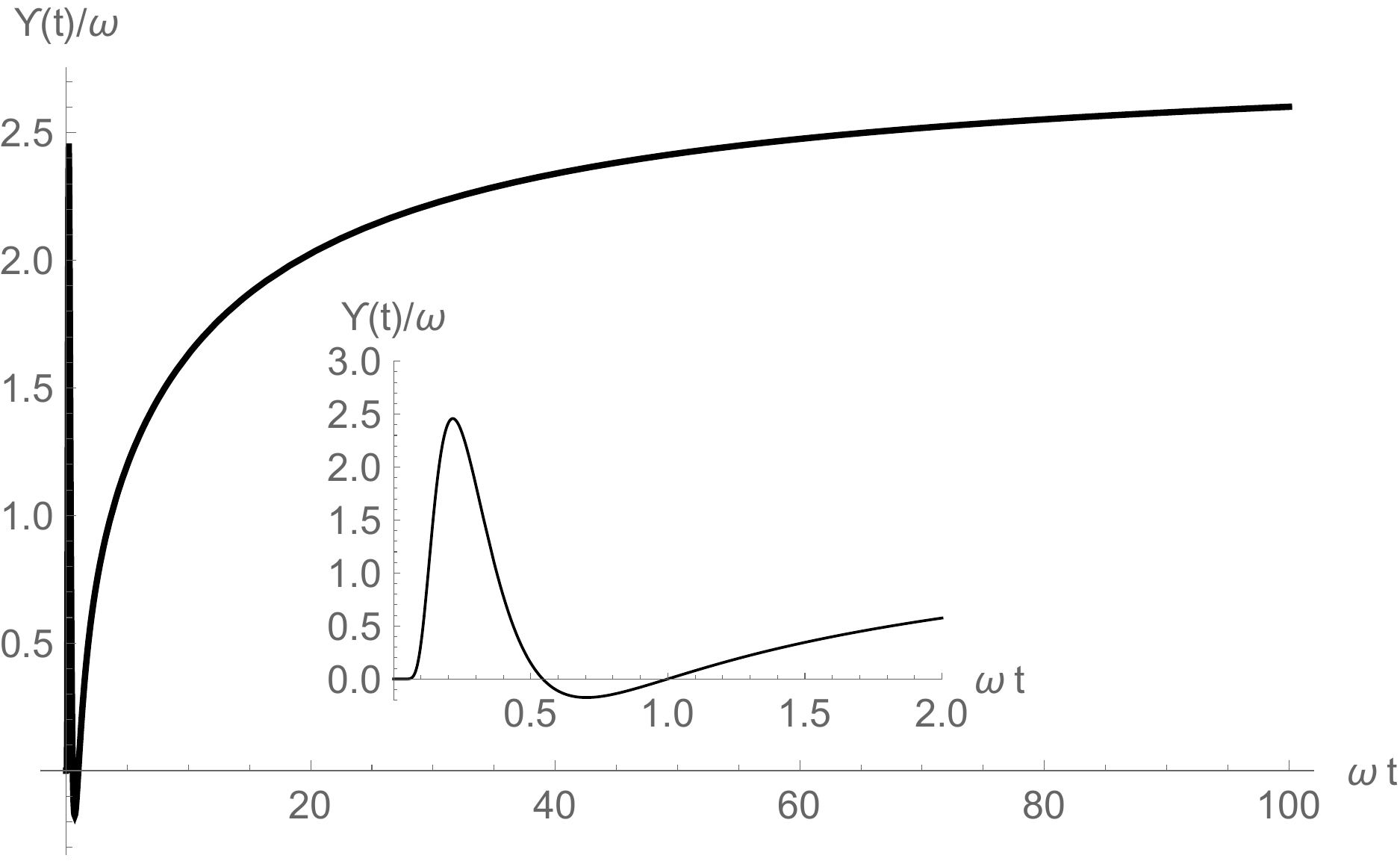}\hspace{.4in}\includegraphics[scale=0.4]{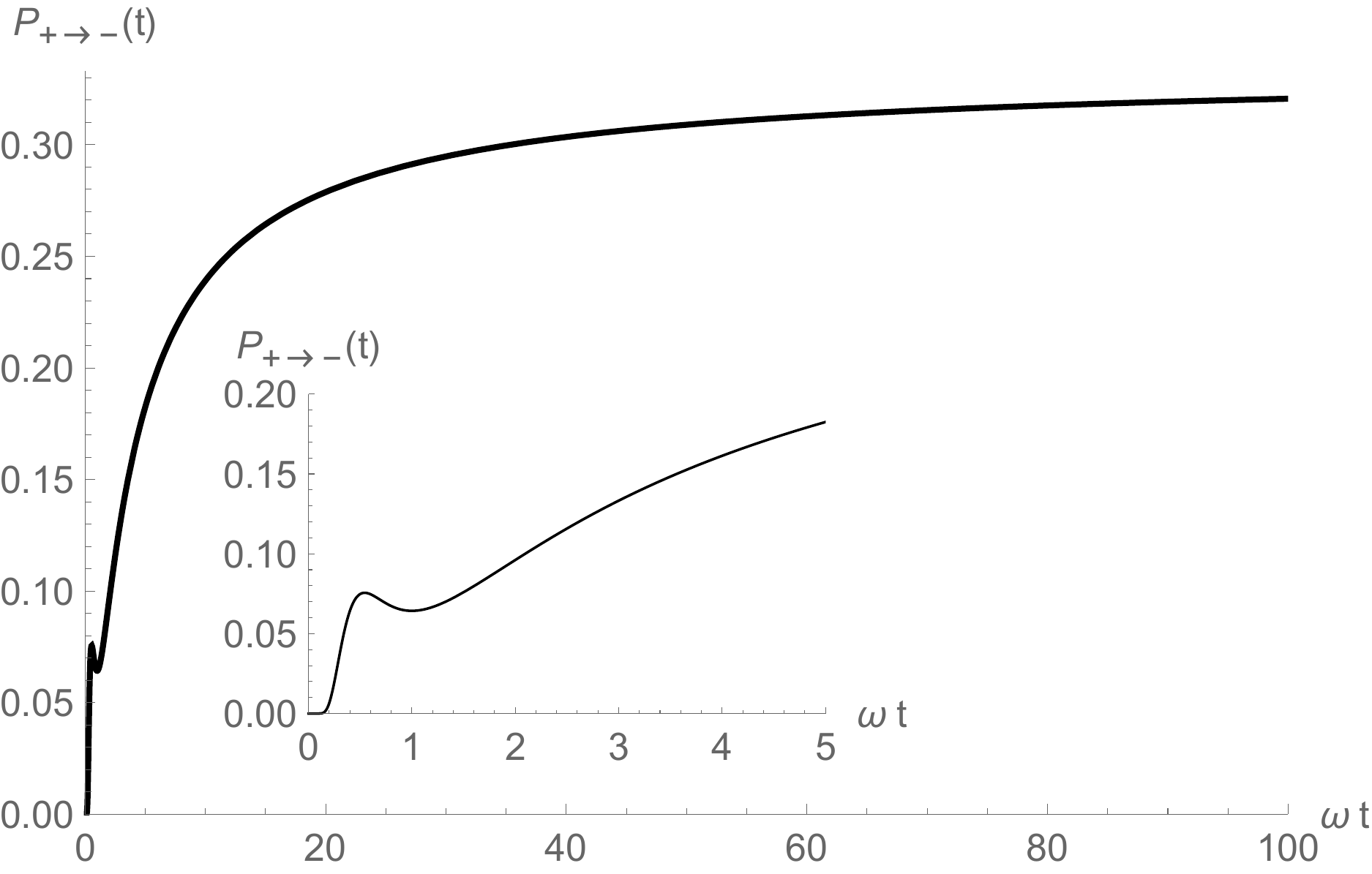}\\
\end{array}$
\caption{Example 2 with parameters $\gamma =1$ and $\Phi(\infty)=\frac{2\sqrt{2}}{3}$. \\
	Left panel: $\Upsilon(t)/\omega$ as a function of $\omega t$ with insert highlighting small $\omega t$ range;\\
	 Right panel: transition probability ${\cal P}_{+\rightarrow -}(t)$ as function of $\omega t$ with insert highlighting small $\omega t$ range.} 
\label{fig2}
	\end{center}
\end{figure*}

The transition probability at time $t$  is 
\bea
{\cal  P}_{+\rightarrow-}(t)&=&\frac{1}{2}-\frac{1}{2}D_2(\omega t)
\eea
which is also plotted as a function of $\omega t$ for $0\le \omega t \le 100$ with the same parameters. Starting at zero when $t=0$, the transition probability initially stays very small, and then sharply rises to a local maximum at $\omega t \simeq0.5$ before dipping to local minimum at $\omega t \simeq 1$. Once again, this behavior is shown in the embedded box in the figure. Subsequently, the probability continues to increase reaching the value 
\bea
{\cal  P}_{+\rightarrow-}(\infty)&=&\frac{1}{2}-\frac{1}{2}\frac{\omega}{\sqrt{\omega^2+\Upsilon^2(\infty)}}|_{\frac{\Upsilon(\infty)}{\omega}=\sqrt{8}}=1/3
\eea
consistent with the general result. Once again, this probability is independent of the Hamiltonian parameter $\gamma \ge 1$ and depends only on the values appearing in $H(\infty)$.\\
\\
Example 3: As a final example, consider
\bea
\Phi(t)&=& \Phi(\infty) (1- e^{-\gamma (\omega t)^2}\cos^2(\beta \omega t))
\eea 
with $0\le  \Phi(\infty) <1$, $0<\gamma \le 1$ and $0\le \beta < 1/4$. Note that $\Phi(0)=\frac{d\Phi(t)}{dt}|_{t=0}=0$ as required while $\frac{d^2\Phi^2(t)}{dt}|_{t=0}=\Phi(\infty)2\omega^2(\gamma +\beta^2)$.\\
This time the function asymptotically approaches the finite value $\Phi(\infty)$ as a Gaussian in $\omega t$ with a modulation in $\cos^2(\beta \omega t)$ when $\beta \ne 0$.

The Hamiltonian term producing such a  factor is 
\bea
\Upsilon(t)&=& \omega \frac{N_3(\omega t)}{D_3(\omega t)}
\eea
where
\bea
N_3(\omega t)&=&\Phi(\infty)\biggr(1-\biggr((1+4\gamma^2\omega^2 t^2-2\gamma)\cos^2(\beta \omega t)\cr
&&~~+4\beta \omega t \sin(2\beta \omega t)-2\beta^2 \cos(2\beta \omega t)\biggr)e^{-\gamma (\omega t)^2}\biggr)
\eea
and 
\bea
D_3(\omega t)&=&\Biggr(1-\Phi^2(\infty)\Biggr(\biggr(1 - \cos^2(\beta \omega t)e^{-\gamma (\omega t)^2}\biggr)^2\cr
	&&~~ +\biggr(2\gamma \omega t \cos^2(\beta \omega t) +\beta \sin(2\beta \omega t) \biggr)^2e^{-2\gamma (\omega t)^2} \Biggr)\Biggr)^{1/2}
\eea
In the figure, we display the ratio $\frac{\Upsilon(t)}{\omega}$ as a function of $\omega t$ for $0\le \omega t \le 4$ with the parameters $\gamma =1, \beta = 0.1$ and $\Phi(\infty)=\frac{2\sqrt{2}}{3}$. In this case, $\Upsilon (t)$ initially falls from its value of $\frac{4\sqrt{2}}{3}\omega$ at $t=0$ to a minimum of around $ -0.4$ at $\omega t \simeq 1.05$ before rising again to approach its asymptotic  limit
\bea
\Upsilon(\infty)&=&\omega \frac{\Phi(\infty)}{\sqrt{1-\Phi^2(\infty)}}=8
\eea
Here the approach to $t\rightarrow \infty$ sets in very rapidly as a Gaussian in $\omega t$. \\

\begin{figure*}
	\begin{center}
		$\begin{array}{cc}
		\includegraphics[scale=0.4]{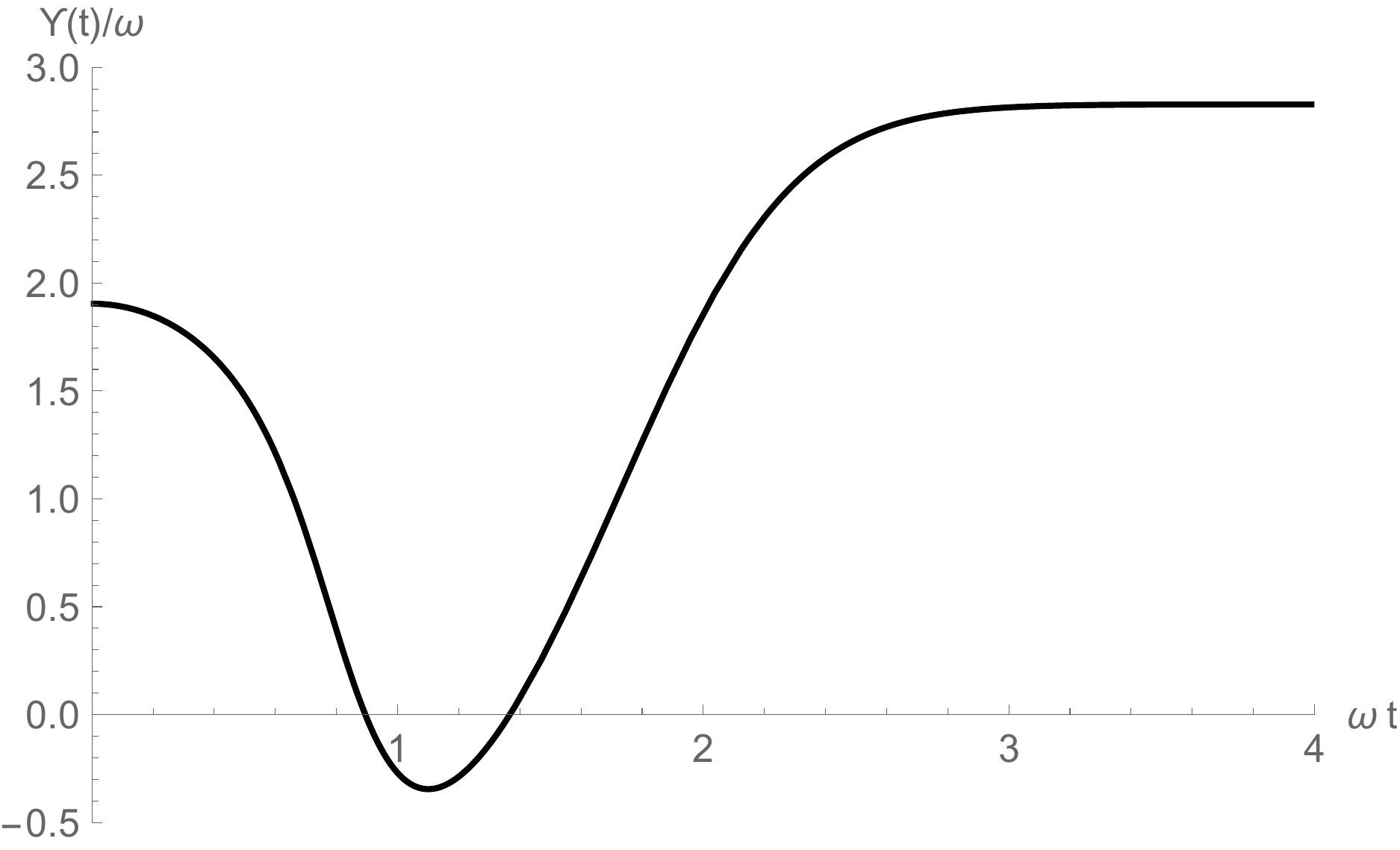}\hspace{.4in}\includegraphics[scale=0.4]{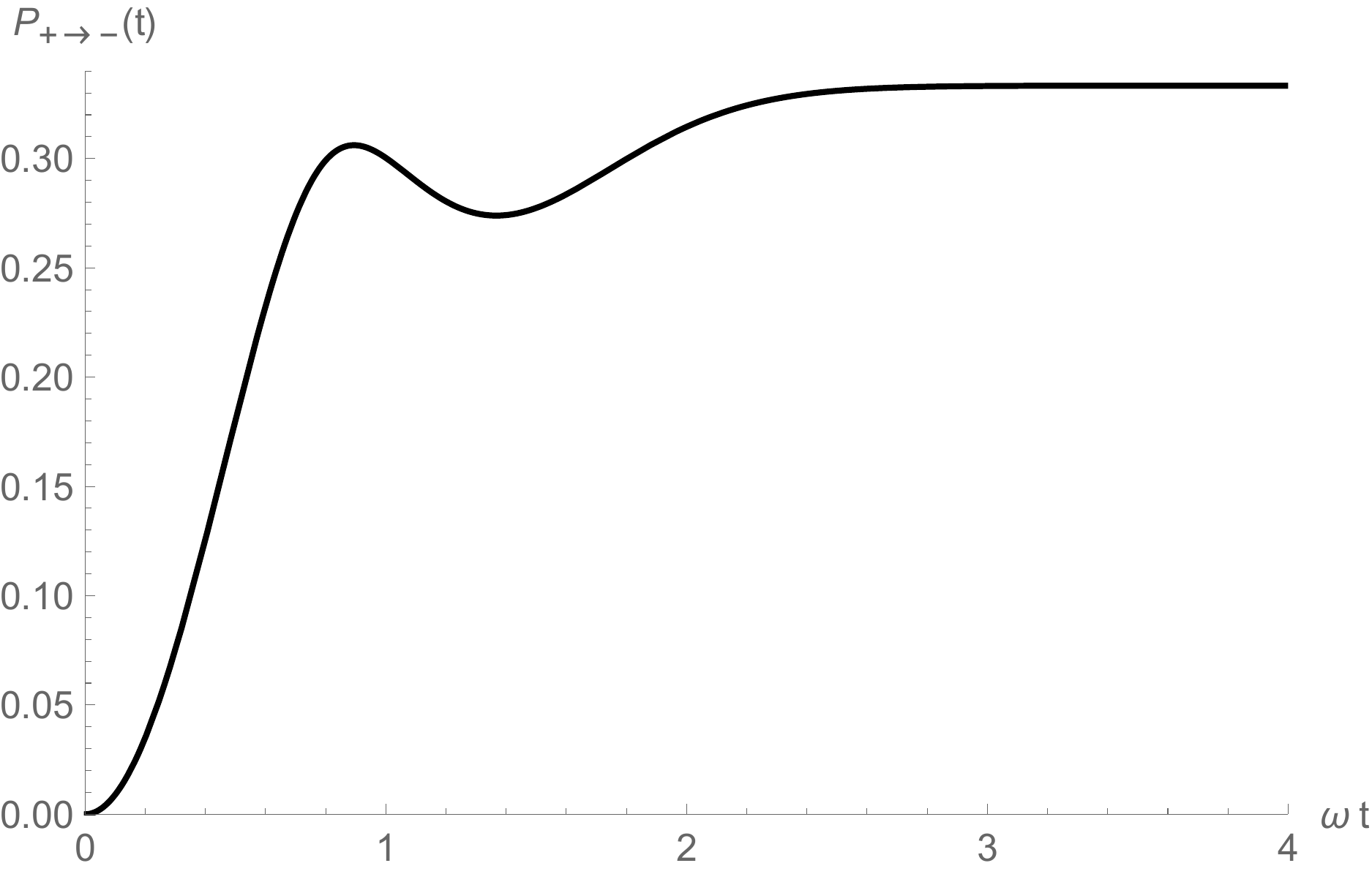}\\
		\end{array}$
		\caption{Example 3 with parameters $\gamma =1~;~\beta=0.1$ and $\Phi(\infty)=\frac{2\sqrt{2}}{3}$. \\
			Left panel: $\Upsilon(t)/\omega$ as a function of $\omega t$;\\
			Right panel: transition probability ${\cal P}_{+\rightarrow -}(t)$ as function of $\omega t$.} 
		\label{fig3}
	\end{center}
\end{figure*}

The resulting transition probability at time $t$ is
\bea
{\cal  P}_{+\rightarrow-}(t)&=&\frac{1}{2}-\frac{1}{2}D_3(\omega t) 
\eea
and is also plotted as a function of $\omega t$ for $0\le \omega t \le 3$ with the same parameters. In this case, the transition probability rises rapidly from zero exhibiting a gentle oscillation for $1 \le \omega t \le 2$ before rapidly settling to its  $t\rightarrow \infty$ limit which is given by the general result
\bea
{\cal  P}_{+\rightarrow-}(\infty)&=&\frac{1}{2}-\frac{1}{2}\frac{\omega}{\sqrt{\omega^2+\Upsilon^2(\infty)}}=1/3
\eea
which once again depends only on the Hamiltonian parameters at $\infty$ and is independent of parameters entering at finite times. \\
\\
We have exhibited three  examples of Hamiltonian driving terms having  quite distinct behaviors at finite times, but each of which producing the same transition probability at $t\rightarrow \infty$.

%\newpage 

\end{document}